\documentclass{icrc29}
\usepackage{graphicx,amssymb,amsmath,times,epsfig}
\begin{document}
\title[Absolute Calibration of the Auger Fluorescence
  Detectors]{Absolute Calibration of the Auger Fluorescence Detectors} 
\author[P. Bauleo et al.]{P. Bauleo, J. Brack, L. Garrard,
        J. Harton, R. Knapik, R. Meyhandan, 
        A.C.~Rovero, \newauthor A. Tamashiro,
       and D. Warner, for the Auger Collaboration}
\presenter{Presenter: A.C. Rovero (rovero@iafe.uba.ar), arg-rovero-AC-abs1-he15-poster}
\maketitle
\begin{abstract}
Absolute calibration of the Pierre Auger Observatory fluorescence
detectors uses a light source at the telescope aperture.  The
technique accounts for the combined effects of all detector components in
a single measurement.  The calibrated 2.5 m diameter light source
fills the aperture, providing uniform illumination to each pixel.
The known flux from the light source and the response of the 
acquisition system give the required calibration for each pixel.  In
the lab, light source uniformity is studied using CCD images and the
intensity is measured relative to NIST-calibrated photodiodes.
Overall uncertainties are presently 12\%, and are dominated by
systematics.
\end{abstract}
\section{Introduction}
The reconstruction of air shower longitudinal profiles and the ability
to determine the total energy of a reconstructed shower depend on
being able to convert ADC counts to a light flux at the aperture for
each channel that receives a portion of the signal from a shower.  To
achieve this objective, it is necessary to have some method for
evaluating the response of each pixel to a given flux of incident
photons from the solid angle covered by that pixel, including effects
of aperture projection, optical filter transmittance, reflection at
optical surfaces, mirror reflectivity, pixel light collection
efficiency and area, cathode quantum efficiency, PMT gain, pre-amp and
amplifier gains, and digital conversion.
\begin{figure}[b]
\begin{center} 
\mbox{\epsfig{figure=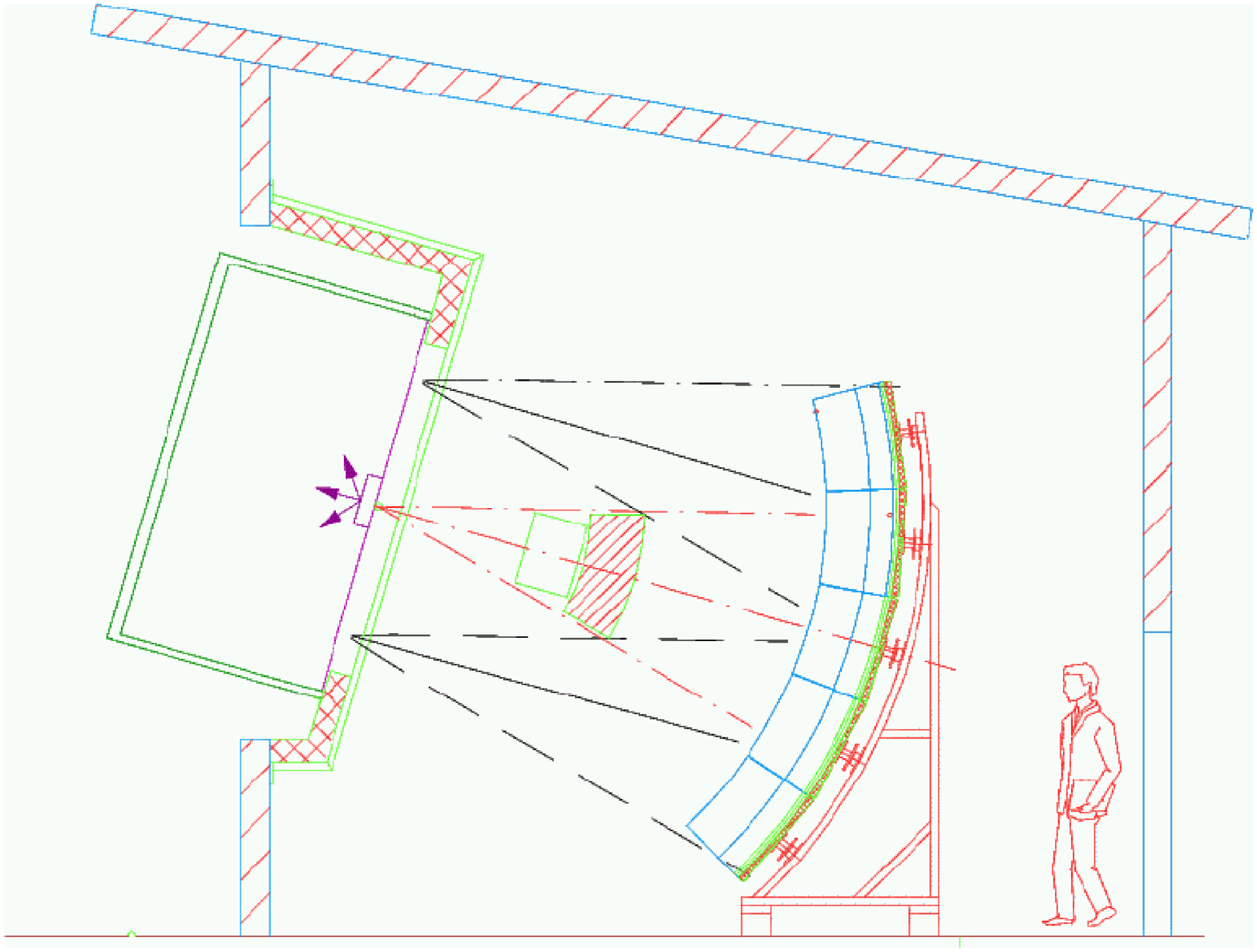,angle=0,height=5cm}
      \epsfig{figure=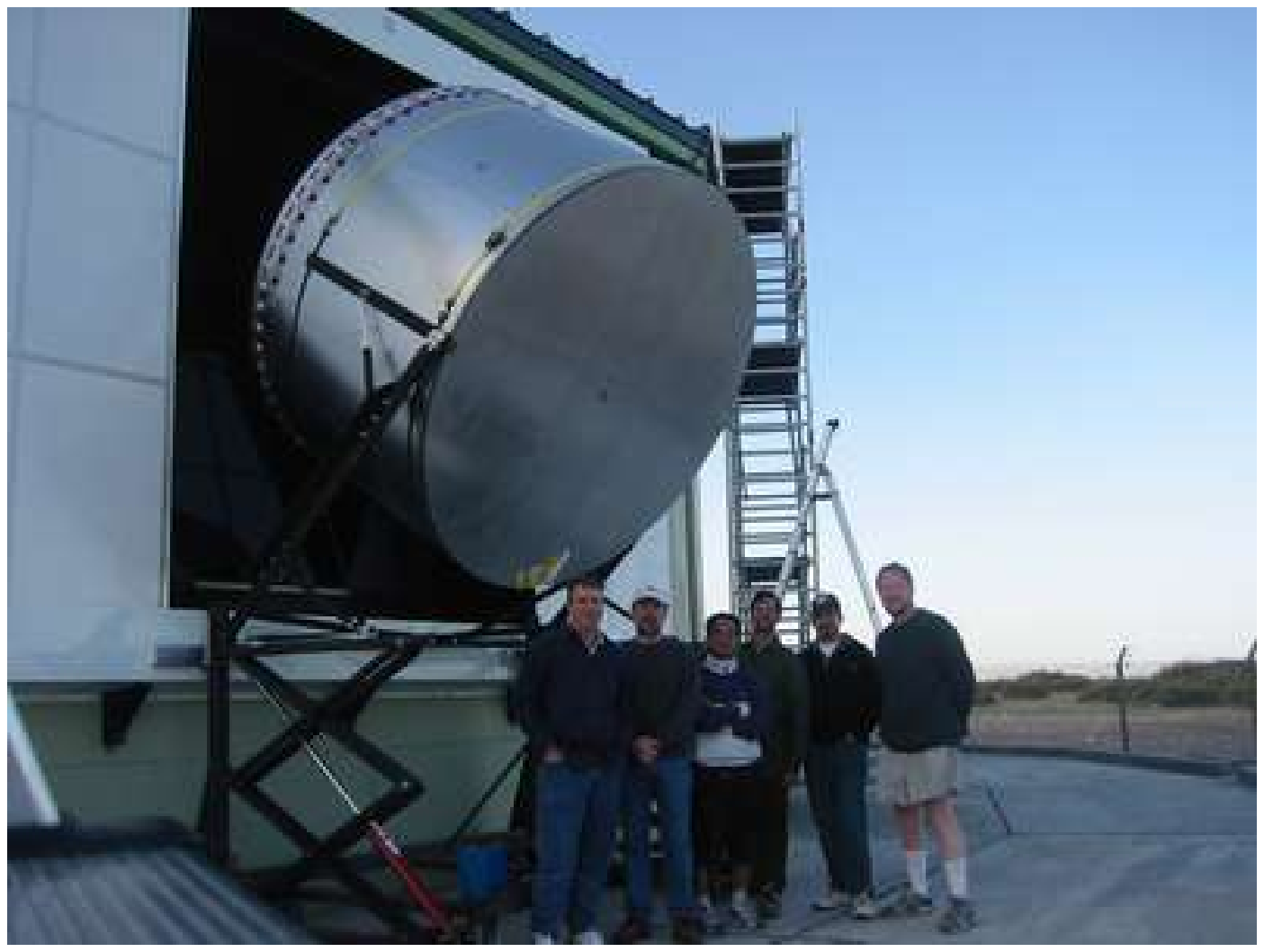,angle=0,height=5cm}}
\end{center}
\caption{A schematic showing the drum mounted in a
  telescope aperture, and a picture of the drum being raised into position.}
\label{schematic-pic}
\end{figure}

While this response could be assembled from independently measured
quantities for each of these effects, the Auger FD calibration group
is using an alternative method in which the cumulative effect is
measured in a single end-to-end calibration.  The technique is based
on a portable 2.5 m diameter drum-shaped light source which mounts on
the exterior of the FD apertures (see Fig.~\ref{schematic-pic}),
providing a pulsed photon flux of 
known intensity and uniformity across the aperture and simultaneously
triggering all the pixels in the camera.

The absolute calibration of the drum light source intensity is based
on a set of UV enhanced Si photodetectors, calibrated at
NIST~\cite{NIST}.  While the small surface area and low response of these
detectors preclude detection of the small photon flux from
the drum surface directly, the NIST calibration can be transferred to
a more sensitive PMT/DAQ system.
Prototype work is described in Ref.~\cite{abscal}; the current
procedure is presented here, including a more direct calibration of
the drum intensity.

Ideally, this calibration would occur at many wavelengths in the N$_2$
spectrum, between 300 and 400 nm, and at several intensities.  At
present, the system is single-wavelength, using UV LEDs emitting in a
narrow band around 375 nm.  Modifications to the
present system are under construction to allow multi-wavelength
calibration, using a Xenon light source, light guide and bandpass
filters.

\section{Light source and drum}
The light source consists of a pair of pulsed UV LEDs~\cite{nichia}
($375\pm12$ nm) mounted coaxially in a 2.5 cm diameter x 2.5 cm long
teflon cylinder.  
A silicon detector attached to
the end of the cylinder monitors the relative light output for each
pulse of the LEDs.  This source is mounted in a 15 cm diameter
reflector cup.

The cup is mounted flush to the center of the drum front surface,
illuminating the interior of the 2.5 m diameter cylindrical drum, 1.4
m deep.  The lightweight drum, shown in Fig.~\ref{schematic-pic}, was
constructed in sections, using 
laminations of honeycomb core and aluminum sheet.  The sides and back
surfaces of the drum interior are lined with Tyvek, a material diffusively
reflective in the UV.  The front face of the drum is a 0.38 mm thick
Teflon sheet, which transmits light diffusively.  

\section{Drum relative uniformity measurements}
Uniformity of light emission from the drum surface is important, since
the pixels in an Auger fluorescence detector camera view the
aperture at varying angles.  In addition, for each pixel, a different
part of the aperture is blocked by the camera itself.  So
studies were made of uniformity of emission across the surface and as
a function of viewing angle.

\begin{figure}[ht]
\begin{center} 
\mbox{\epsfig{figure=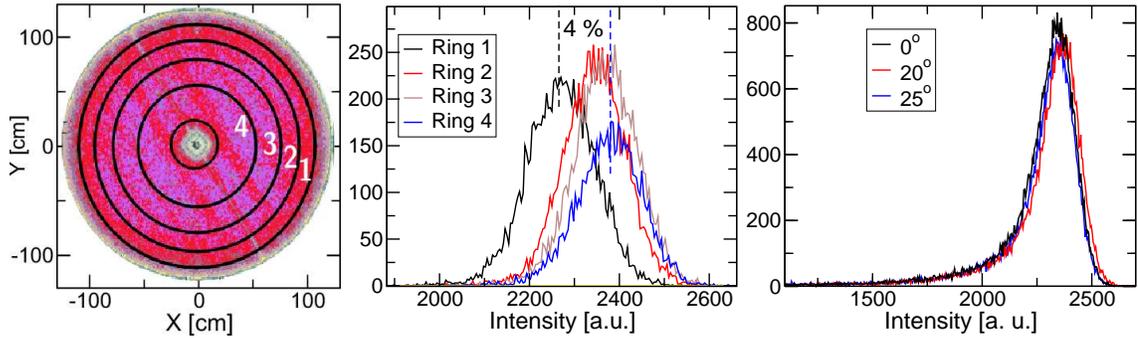,height=4.4cm}
      \epsfig{figure=figure2b.eps,angle=0,height=4.4cm}
      \epsfig{figure=figure2c.eps,angle=0,height=4.4cm}}
\end{center}
\caption{From left to right:
Color-enhanced CCD image at 0$^0$ drum angle, showing the
defined rings for relative intensity analysis.
Production deformations in the teflon material can be seen;
A plot of the observed pixel intensities in the defined
regions of the drum, shown in the previous
figure;
The results of angular relative intensity measurements at 0, 20,
and 25$^0$. 
} 
\label{drum-unif}
\end{figure}

These uniformity measurements were made using a CCD, viewing the 
emitting surface of the drum from a distance of 14 m (see
Fig.~\ref{drum-unif}).  Images were recorded with
the drum axis at angles of 0, 10, 20 and 25 degrees relative to the
CCD axis, covering the range of the Auger telescope field of view 
(0--21$^0$).  For these images, the UV LEDs were powered continuously.

The 0$^0$ image was used to analyze the uniformity of emission over
the drum surface.  Using software,
concentric circles were drawn on the image, defining annular regions
of increasing radius, as shown in the figure.  
The intensity of the pixels in each region was
analyzed to obtain the intensity as a function of radius.  
In the area defined by the 2.2 m aperture
radius, the relative uniformity of intensity is constant over the area
to about $\pm$2\%.
The variation with viewing angle of a section of the drum image is
also shown in
Fig.~\ref{drum-unif}.  

The fact that the drum intensity decreases at larger radii and that the
area viewed decreases as cos$\theta$ with viewing angle is partly
compensated for by the location of the camera shadow for pixels in
different parts of the camera.  Central pixels view the fuller 0$^0$
area but the more intense center of the drum is shadowed; large angle
pixels and those in the corners of the camera see a reduced area, but
the camera shadow moves off-center, exposing more of the brighter drum
center.

The small drum non-uniformities that are measured 
(diagonal stripes in Fig.~\ref{drum-unif}, 
intensity decrease with radius, etc.)
indicate that the FD pixels
see similar intensities integrated over the drum surface.  While
perfect drum uniformity is desirable, non--uniformities which are
small and well mapped over the surface of the drum are acceptable.  A
ray-tracing program using the uniformity and angular intensity
information from the CCD images shows less than 1\% variation in total
flux seen by the pixels, and corrections are applied for these
variations. 

\section{Drum calibration}
To establish the absolute flux of photons emitted from the drum
surface, a reference PMT is placed on the drum axis, 14 m from the
surface.  The LED light source in the drum is pulsed for a series of
5 $\mu$s pulses and the charge from the PMT is integrated and recorded
for each pulse, resulting in a histogram of the distribution of the
observed integrated flux.

On an optical bench, the PMT is then exposed to a small diffuse LED
light source of adjustable intensity similar to that of the drum.  The
LEDs are pulsed with the same driving circuitry as for the drum, and
the intensity is set to a series of values producing a series of
histograms with centroids surrounding that from the drum measurement,
above.  At each of these intensity settings a second measurement is
made in which the PMT is replaced by the NIST-calibrated photodiode
and a neutral density filter in the source is removed, increasing the
intensity to a level measurable by the photodiode.  For this second
measurement, the LEDs are run in DC mode.  The relationship
of PMT response to photodiode current is found to be very linear.  The
flux of photons at 14 m from the drum surface is then calculable from
the active area of the photodiode, the neutral density filter
reduction factor, the LED pulsed/DC duty factor, 
the NIST calibration for the photodiode, and the
value of photodiode current corresponding to the PMT-drum centroid, as
interpolated from the linear response-current relationship above.  

\section{Calibration Results}
\begin{figure}[ht]
\begin{center} 
\mbox{\epsfig{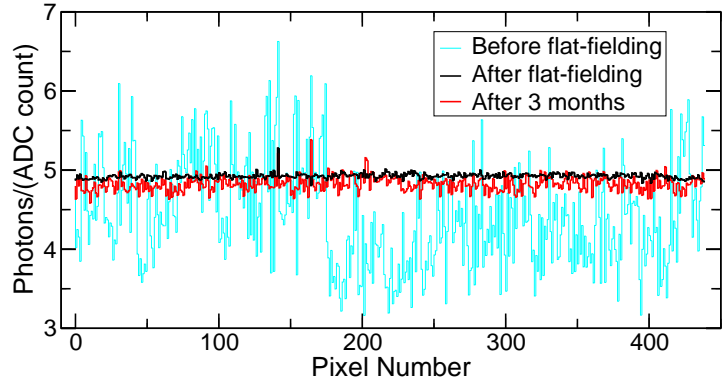}}
\end{center}
\caption{The effects of drum calibration on the 440 pixels of the
  camera in bay 5 at Los Leones.   
  Blue: as commissioned;
  Black: immediately after calibration; Red: typical variation seen
  during next calibration.
}
\label{flat}
\end{figure}

Use of the drum for gain adjustment and calibration provides a known,
uniform response for each pixel in a detector as shown for one camera
in Fig.~\ref{flat}.  The typical sensitivity for Auger FD pixels is
approximately 5 photons per ADC count.  At present, the absolute
calibration is performed every three months.  
The long-term variations of camera response are typically within
2\% of the calibration and are associated with several
factors, including gain shifts due to PMT
exposure to varying night sky intensities during the previous few nights
of data taking.
These small variations in 
pixel response are tracked several times per night by
a relative calibration system, and appropriate corrections for
each pixel are entered daily in the analysis database (see associated
Auger paper in this ICRC~\cite{antonio}).
The absolute calibration as described here has been checked for some
pixels using remote laser shots at 337 and 355 nm~\cite{laser-calib}.

Several effects on the calibration have been studied, including:
\begin{itemize}
\item 
Reflections from detector components back to the drum emitting
  surface when the drum is mounted in the aperture, making the surface
  brighter than when calibrated in the lab.  This is dominated
  by reflections from the UV filter back to the teflon face of the
  drum.  This is a 4\% effect, with a $\pm$1\% uncertainty. 
\item 
Temperature variations between calibration lab and FD apertures.  LED
  intensity, electronics, and monitoring equipment are all temperature
  sensitive.  Varying effect; 3\% uncertainty. 
\item 
Shape of LED light emission pulse.  Pulsed LED duty factor compared to
  measured DC intensity is a component of the calibration
  calculation.  This is a 2\% effect; 1\% uncertainty. 
\end{itemize}
While the general technique of Auger FD calibration is now well
established, the equipment and detailed procedures are still evolving.
Recent
improvements of the lab optical equipment will improve the overall
uncertainties in the next calibration.  At present, uncertainties
total 12\% and are dominated by contributions associated with the
optical setup in the lab (10\%) and a conservative estimate of the
nightly variations (5\%).  Uncertainties of about 8\% are achievable.


\begin{thebibliography}{99}

\bibitem{NIST} National Institute of Standards and Technology,
   U.S. Dept. of Commerce,  Calibration Program, Gaithersburg, MD
   20899-2330;  NIST Special Publication 250-41, 1998.

\bibitem{abscal} J. Brack, R. Meyhandan, G. Hofman, J. Matthews,
  Astroparticle Physics 20, 653 (2004). 

\bibitem{nichia} Nichia America Corp., NSHU550 UV LED

\bibitem{antonio} C. Aramo et al., ICRC 2005, ita-insolia-A-abs1-he15-poster.

\bibitem{laser-calib} Auger technical note GAP-2002-10, 
 B. Dawson, B. Fick, J. Matthews, M. Roberts, P. Sommers,
 http://www.auger.org/admin/GAP\_Notes/GAP2002/gap2002-010.pdf,

\end{thebibliography}
\end{document}